%% file: Main.tex
\documentclass[runningheads]{llncs}
\usepackage{graphicx}
\usepackage{color}
\usepackage{enumitem}

\begin{document}
\title{\textit{Label it be!} A large-scale study of issue labeling in modern open-source repositories}
\titlerunning{\textit{Label it be!} A large-scale study of issue labeling in modern OS repositories}
%
\author{Joselito Júnior \and 
        Gláucya Boechat \and
        Ivan Machado}
\authorrunning{J. Júnior et al.}
%
\institute{Federal University of Bahia - UFBA, Salvador - Bahia, Brazil \and
\email{\{joselito.mota, glaucya.boechat, ivan.machado\}@ufba.br}}

\maketitle              

\input{Sections/0_Abstract}
\input{Sections/1_Introduction}
\input{Sections/2_Background}
\input{Sections/3_Methodology}

\input{Sections/4_Results}
\input{Sections/5_Discussions}

\input{Sections/6_Threats_Validity}

\input{Sections/7_Related_Work}
\input{Sections/8_Conclusions}

\bibliographystyle{splncs04}
\bibliography{references}

\end{document}

%% file: Sections/0_Abstract.tex
\begin{abstract}

In a wave of growth, open-source projects need to modernize and change how they deal with processes, methods, and communication with their contributors. We could observe that open-source projects are constantly evolving to improve their management of the entire community.  Starting with community communication, software development, managing open-source projects faces crucial challenges. One of the enabling environments that open-source communities found to achieve community communications objectives was code repositories with integration with issue trackers. Using issue trackers in their projects should encompass an infrastructure capable of hosting the project source code and community participation. Some issue trackers use a structure in which the issue's title and description are the key information. However, we have observed a slight change in this strategy over the years, as more and more data are fundamental to solving the issue. For example, labeling the issues could enable users to provide the issue with more contextual information. By understanding how modern issue trackers handle issue labeling, this study analyzes the impact, engagement, and influence that labels have on the Github repositories, based on a database of 10,673,459 issues mined from 13,280 repositories in 180 Github featured topics. We found that 78.75\% of the repositories label their issues, with more adherence from those repositories that contain big numbers of issues. The labeling practice is essential and prioritized as a first step in the issue resolution process in 65.91\% of the first events. Issues with labels draw more attention and impact by collecting more subscribers, assigns, and comments, helping to engage contributors to the resolution.

\keywords{Github \and Issue labeling \and Issue trackers}

\end{abstract}

%% file: Sections/1_Introduction.tex
\section{Introduction}

Over the years, software development has grown both in size and importance. To accommodate the urging and increasing demand for more robust, reliable, and efficient software solutions rapidly, it is often necessary to join efforts from people based in different locations to deliver their potential to contribute to the software project. Such a scenario is widespread in open-source software projects, where people could work as contributors regardless of their physical location, but rather for their motivations or interests \cite{Hertel:JournalResearchPolicy:2003}.

In this wave of growth, open-source projects need to modernize and change how they deal with processes, methods, and communication with their contributors. That is why there is a constant evolution in software technology and the community and their interactions \cite{Nakakoji:IWPSE:2002}. Open-source projects are constantly changing to improve their management of the community as a whole. Starting with community communication, software development, managing open-source projects stands crucial challenges. Communication within the community and an environment that favors continuous integration and maintenance of systems are first-class dimensions that characterize a representative, and well-structured repository \cite{Munaiah:JournalEmpiricalSoftwareEng:2017} \cite{Rodriguez:RAISE:2012}. One of the enabling environments that open-source communities found to achieve these goals was code repositories with integration with issue trackers. Using issue trackers in their projects should encompass an infrastructure capable of hosting the project source code and enabling the community participation (e.g., developers, testers) to provide detailed information about their systems using the issue-reporting mechanism. The issues would be later on assigned to any community member capable of handling it \cite{Anvik:OOPSLA:2005}.


The use of an issue tracker with the reporting of problems and establishing a problem-solving process brings an essential ally in the maintenance and evolution of the software \cite{Koponen:IFIP:2006}. However, an issue should be described in the most comprehensive way possible \cite{Lientz:JournalACM:1978}, so that it does not make it difficult for developers to understand the problem he should handle \cite{Aranda:ICSE:2009} \cite{Bettenburg:OOPSLA:2007} \cite{Bettenburg:SGISOFT:2008}. Incomplete information and unsuitable classifications can confuse developers about the causes of the problems, leading to correction and validation delays, harming all users who need changes in the software.


Most issue trackers use a structure encompassing the issues' fields of title and description, and these are the key elements an issue should contain. In addition, some trackers have enabled users to attach labels to incorporate more contextual information into the issue. This practice consists of a text field where the user could assign words or small definitions to associate with a given issue. Issue labeling consists of incorporating issues with meaningful text fragments, or simply labels (or tags), that could readily assist software developers in the rapid and objective perception of the reported problem  \cite{Storey:JournalIEEETransactions:2009} \cite{Treude:JournalIEEETransactions:2019}.



Issue labeling is present in Github-based repositories. Therefore, this study seeks to analyze the impact, engagement, and influence that labels have on the Github repositories, based on a database of 10,673,459 mined issues from 13,280 repositories in 180 Github featured topics. 


In this study, we addressed the following Research Questions (RQ):

\begin{enumerate}[label=\bf RQ\arabic*.,leftmargin=1.6cm]
    \item \textbf{How representative is the labeling activity practice in the repositories?} This RQ aims to analyze and show how representative the labeling activity is in repositories by applying labeling issues in all the repositories from the database. We want to study the adherence and popularity of issue labeling by project contributors to discover any usability on the repository.
    \item \textbf{What is the impact of labels on issue engagement that differs from non-labeled issues?} Acknowledging that all issue ``movements'' are traceable, this RQ investigates the impacts of the labeling activity in issues by the contributors' engagement. We studied certain events as users' subscriptions to follow the discussion in the issue, the assignment event to a contributor in both labeled and unlabeled issues. We looked at the advantages and disadvantages of label issues from the perspective of repository improvement.

    \item \textbf{What is the level of priority project contributors give the labeling activity when opening a new issue?}
    In this RQ, we took the perspective of the priority given by the project for labeling when reporting an issue. We study when the labeling in the issue's event list occurs to understand if this is a first recurring practice and is prioritized in the first moments in the resolution process through repositories.

    \item \textbf{What is the relevance of labeling issues for repositories with a different number of issues?} This RQ aims to analyze the labeling activity in repositories with a different number of issues. The analyzed repositories were split into four groups to reach this goal, depending on the number of issues each has. It is possible to make inferences that could help understand and compare the labeling practice's recurrence in different repositories sizes by analyzing such a perspective.

\end{enumerate}

%% file: Sections/2_Background.tex
\section{Background}

\subsection{Repositories}

Repositories have different ways to work, and as such, but the literature encompasses many practices and behaviors that characterize a repository. Munaiah et al.  \cite{Munaiah:JournalEmpiricalSoftwareEng:2017} described that a repository's representativeness involves seven dimensions: community, continuous integration, documentation, history, issues, license, and unit testing. 

\textcolor{black}{According to Rodriguez et al. \cite{Rodriguez:RAISE:2012}, Software Engineering repositories have concepts as:} source code, software configuration management (SCM), issue tracking, developers' messages, user messages, project metadata, and usage data. Those concepts involve elements that projects must have to frame in well-structured repositories. 

It is worth mentioning that either the two definitions above address issue tracking (or bug repositories) and issues as essential factors to define a repository. Bug repositories or issue trackers are used to report and maintain problems or suggestions about the software \cite{Anvik:ICSE:2006}. The importance of a well-structured organization and adequate communication is also essential for solving these problems or even avoid introducing new defects \cite{Bernardi:ECSMR:2012} \cite{Mockus:SIGSOFT:2010}. These reporting channels allow the community to help in the software maintenance process. The combination of user feedback, community work, and maintainability are critical factors for the success of repositories \cite{Schneider:ICSE:2003}, including in Github repositories  \cite{Dabbish:CSCW:2012}.


\subsection{Issues}

The issue as a defect follows a timeline that consists of insertion, detection, and removal. It is called the defect life-cycle \cite{Zubrow:IEEEStandards:2009}. \textcolor{black}{Users commonly report the detection of a defect through issue reports.} Based on the reports' detailed information, developers try to reproduce the defect as described in the issue to understand its likely causes and then proceed to the resolution phase.

The components present in Github issues are very similar to those of known issue trackers, \textcolor{black}{but those components} has some features that other platforms do not have. When it is necessary to write an issue, it is required to fill in some components such as title, description, and comments, encompassing all necessary details to enable reproducing the issue. In the description and comments fields, it is possible to add lists, mark code snippets, quotes, warnings, and other elements through the Markdown syntax \cite{Github:Site:2020}.

Github also includes other fields that contributors could use in a Github issue besides the title, description, and comments. A Github issue has events that show all the activities performed in the issue. It also enables to configure sending notifications to users about updates in the discussion. Besides, it includes a feature to assign issues to contributors \cite{Github:SiteAssigningIssues:2020}, the issue status informing whether the issue status, i.e., either \textit{open} or \textit{closed}, and also, the focus of this study, the labeling issues activity.

\subsection{Labels}

A label is one type of issue component and one more option that Github provides developers who wish to add extra information about the reported issue. The platform provides a standard list of nine labels with a short description for each. They are: \textit{bug, documentation, duplicate, enhancement, good first issue, help wanted, invalid, question} and \textit{wontfix}. Developers responsible for the repository could also create custom labels that better suit their needs.

The concept of tags in a social system is a free field for text insertion and personalization \cite{Storey:JournalIEEETransactions:2009}. The Github labels follow a similar idea for those tags and can add accessible and quick information about the issue.

Labels have different characteristics, whether created by either humans or machines \cite{Nayebi:JournalIEEESoftware:2018}. They could apply to any related function, including reuse, management, and re-finding information \cite{Storey:JournalIEEETransactions:2009} in many software artifacts such as architecture, components, documentation, testing, and others \cite{Treude:JournalIEEETransactions:2019}. For an issue, labeling refers to bugs and improvements \cite{Bissyande:ISSRE:2013}.

%% file: Sections/3_Methodology.tex
\section{Methodology}

\subsection{Searching for relevant repositories}

In this study, we looked for repositories with many events or that are influential. Github hosts a page \cite{Github:Site:2020} with all featured topics on the platform, which means a classification of what types the repositories are. Altogether there are 180 topics from different subjects and domains, such as programming languages, browser plugins, compilers, and other topics. Each topic has thousands of repositories. Given that \textbf{number of stars} is one the most widely used popularity metrics of a Github repository \cite{BORGES2018112} we used such a counting to select the most popular ones.

\subsection{Mining repositories issues and preprocessing issues' components} 

We mined data about issues and pull-requests, which contains the following metadata: \textit{id, author, title, body, description, status, creation date, events, repository labels, issue labels, reactions}, and \textit{comments}.  The mining process retrieved 10,673,459 issues from 13,280 repositories of Github featured topics from September 2019 to August 2020. All issues were saved and structured in the MongoDB database. The pre-processing phase carried out the process of filtering unnecessary data for our analysis. We used a Python script with regular expressions \cite{regular_expression_Python} to remove unnecessary items such as \textit{code quote, tags, hyperlinks, tables and references, citations and comments repeated replies, warnings, exceptions, class names, paths, special characters, numbers, multiples blank spaces and stop words removal to title, body,} and \textit{comments}. Github uses the Markdown markup language to format its text structures, including all the issue texts \cite{Github:SiteMarkdown:2020}. The usage of Markdown facilitates removing structures that would not add information in the process of analyzing issues. In addition, aspects of its syntax structure have been removed from the issue texts, opening up a corpus text only. An event list records all the issue movements. Understanding this, we examined the list of events to find new evidence of labeling activity through a built-in script.

\subsection{Data analysis}

To conduct the study and answer the \textbf{RQ1}, we perform a search to find labeled issues in the repository with a Python algorithm. If the repository labeled at least one issue with a valid label, the repository performed the labeling. All the activities performed in an issue, such as following a discussion, assignment, closing, labeling, and others, are recorded in a timeline called \textit{events}. The \textbf{RQ2} involved studying the events in all issues of the repositories by searching occurrences of events, such as issue subscription, assignment, and comments of labeled and unlabeled issues. The answer to the \textbf{RQ3} goes through the algorithm analysis of the issue event list to count if the first activity performed on the issue is the labeling activity. The final amount of first labeled issues is analyzed. If the repository does not practice labeling first, then it is analyzed at which stage of the issue the labeling in the repository is mainly used. 

\begin{figure}
\includegraphics[width=\textwidth]{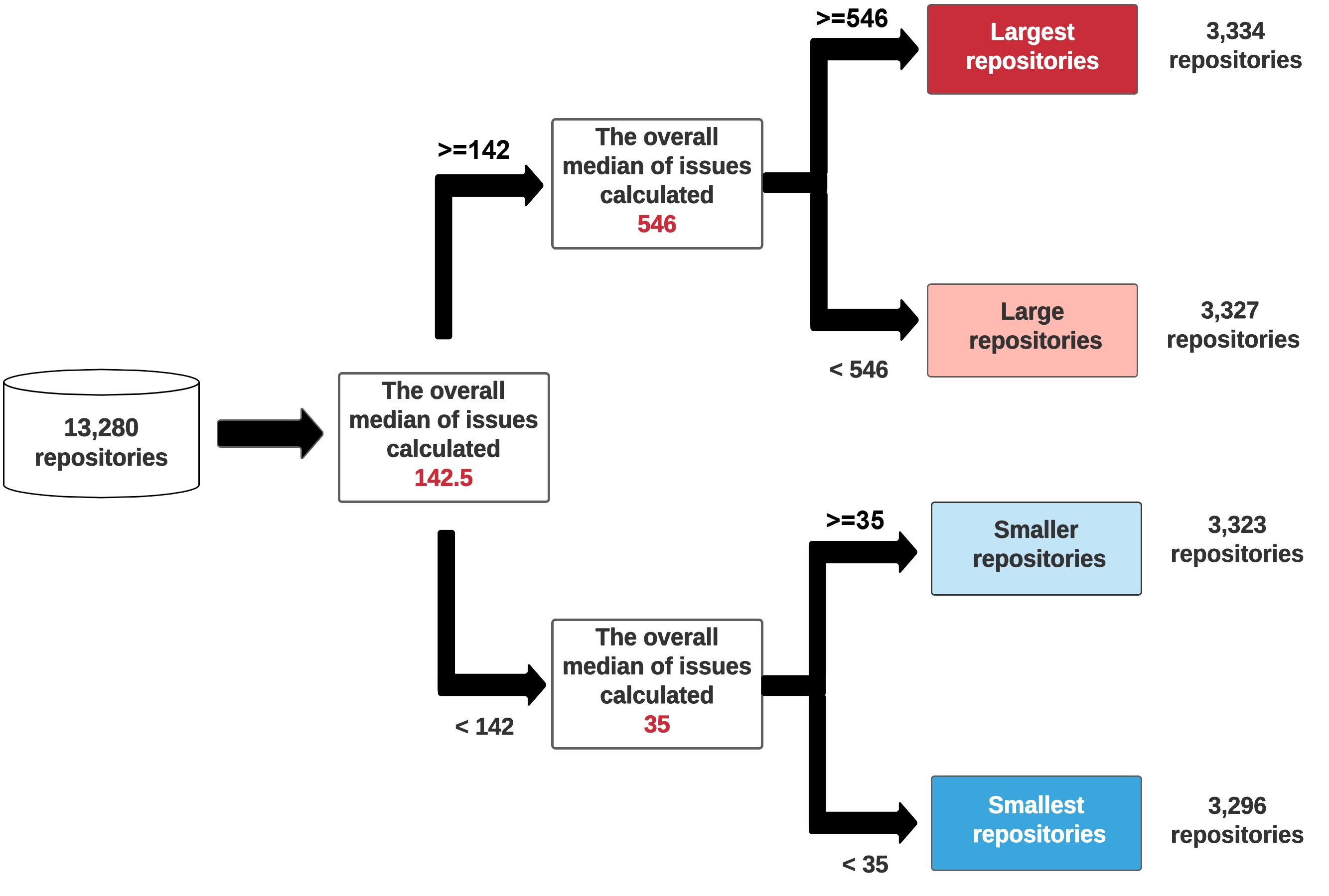}
\caption{Classification process of the size of the repositories by the number of issues} \label{fig_repository_size_groups}
\end{figure}

We split the database into four groups by considering the median amount of issues: \textit{The largest, Large, Small}, and \textit{The smallest}. The \textbf{RQ4} results are reached after the correlation of labeling activities and the number of issues in those groups, and the algorithm rated the repository by the median of issues, as Figure \ref{fig_repository_size_groups} shows. 

The first split used the median number of issues for the entire base and had two parts: repositories with a median greater than or equal and less than 142. Repositories greater than or equal to 142 issues fall into the groups \textit{The Largest} or \textit{Large} repositories. Those with less than 142 issues fall into the groups \textit{Smaller} or \textit{The Smallest} repositories. 

To define \textit{The Largest} and \textit{Large} repositories, we performed another median calculation in the new group. Repositories with a median greater than or equal to 546 issues fall into \textit{The Largest} group, and repositories with less than 546 issues fall into the \textit{Large} group. Similarly, to define either \textit{Small} and \textit{The Smallest} groups, if the median contains a median greater than or equal to 35 issues, it falls into the former, otherwise into the latter one. Figure \ref{fig_repository_size_groups} shows all the group's calculation process. 

%% file: Sections/4_Results.tex
\section{Results}

This section shows all the data results after mining, pre-processing, counting, and analyzing 10,673,459 issues from 13,280 repositories from Github feature topics. The median number of issues in the database was 143, and the standard deviation of 2.889. It is worth mentioning that all the research package, including the database, the results tables, and the scripts implemented and used, are available at \cite{Dropbox:SupplementaryMateria:2021}.

\subsection{\textbf{How representative is the labeling activity practice in the repositories? (RQ1)}}

\begin{figure}
\includegraphics[width=\textwidth]{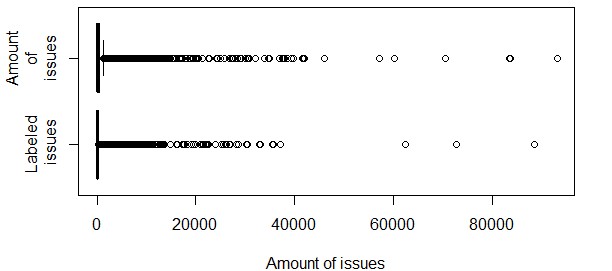}
\caption{Boxplot of the total number of issues with the issues labeled.} \label{fig_boxplot_labeled_issues}
\end{figure}

From the set of analyzed 13,280 repositories, we observed that 10,458 (78,75\%) labeled their issues, while 2,822 (21,25\%) did not label any issues. Although the repository's adherence to labeling is representative, the number of labeled and unlabeled issues is very close. Out of the 10,673,459 issues in the database, 4,917,544 (46,07\%) were labeled, leaving a margin of 5,755,915 (53,93\%) without labeling. The median of issues labeled by the repository is 21, and the standard deviation found was 1,999. Figure \ref{fig_boxplot_labeled_issues} shows the number of issues in the database and the labeled issues.

We found that labeling is somewhat representative, and most repositories frequently use this practice, and labeling occurs almost half the time an issue is reported. However, there is still a considerable number of issues that are not labeled. Each repository has variations in the number of issue labels, so one repository labels many issues, but others label just a few issues. This scenario reflects how each repository handles labeling priority.

\subsection{\textbf{What is the impact of labels on issue engagement that differs from non-labeled issues? (RQ2)}}

Performing a search in the events of all repositories retrieved interesting results about labeled and unlabeled issues. Unlabeled issues have 66.64\% (107,574,958 events) of the events; the labeled issues only have 33.35\% (53,830,132 events). We expected to reach such a number due to a slightly larger number of unlabeled issues. We could find some trends about issue subscribing more deeply by analyzing the events: there were 66,48\% (10,127,463 subscribers) in labeled issues and 33.52\% (5,106,845 subscribers) of unlabeled issues. Issues with labels have more assignments to contributors than unlabeled, 81.72\% (1,736,294 assigns) in labeled issue and 18.27\% (388,224 assigns) in unlabeled issues. We found 57.69\% (20,553,963 comments) of the comments in labeled issues and 42.31\% (15,075,414 comments) when the issue does not have a label.

Although issues without labels have a more significant number of associated events, labeled issues draw more attention and impact repositories issues by collecting more subscribers, assignments, and comments than unlabeled issues, presenting as a factor to consider when using repositories.



\subsection{\textbf{What is the level of priority project contributors give the labeling activity when opening a new issue? (RQ3)}}

\begin{figure}
\includegraphics[width=\textwidth]{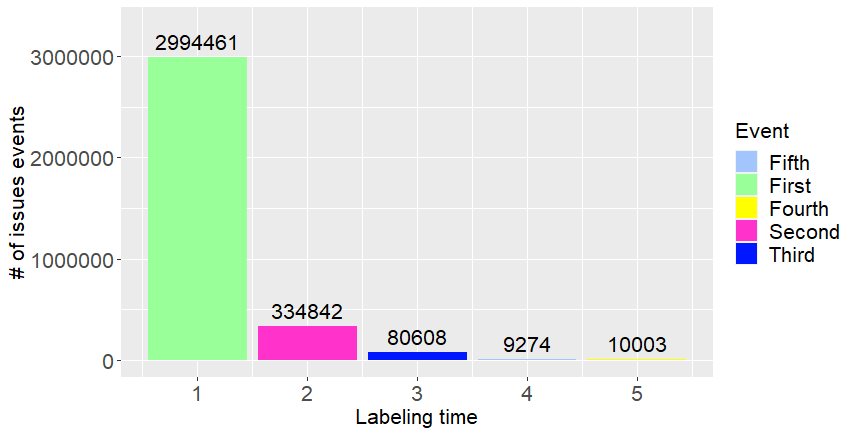}
\caption{When the labeling event often occurs in labeled issues.} \label{fig_labeled_event_moment}
\end{figure}

In this RQ, we analyzed the number of labeled issues. Finding the time for labeling issues is the critical key to know the priority given to this activity. To discover the labeling moment, we accessed the list of events of all issues labeled from each repository in our database and counted all the labeling moments. The events of an issue present some activities carried out, such as assignments, closings, labeling time, and others. Then, we calculate when issues are labeled in first events, the number of issues labeled first when the labeling activity is more frequent, and the frequency it happens. Figure \ref{fig_labeled_event_moment} shows when the labeling event occurs most when an issue is registered.

As Figure \ref{fig_labeled_event_moment} shows, the labeling activity happens massively in the first issue event. 2,994,461 issues were firstly labeled; this represents 60.89\% of the total issues labeled amount. Labeling as the first activity is also dominant in 8,754 (65.91\%) of the repositories. 

From these results, we could infer that the labeling activity, when performed, is prioritized as one of the first events when reporting an issue among several event actions. As a first event prioritized by project contributors many times in labeled issues, we can infer an essential role in the issue resolution process.

\subsection{\textbf{What is the relevance of labeling issues for repositories with a different number of issues? (RQ4)}}

We first analyzed whether the issue tracker size correlates with the labeling activity by performing the Spearman's rank correlation in each repository issue and the ratio of labeled issues. The Spearman's rho value is 0.42, indicating a moderate correlation between repository issue size and labeling activity. 

\begin{figure}
\includegraphics[width=\textwidth]{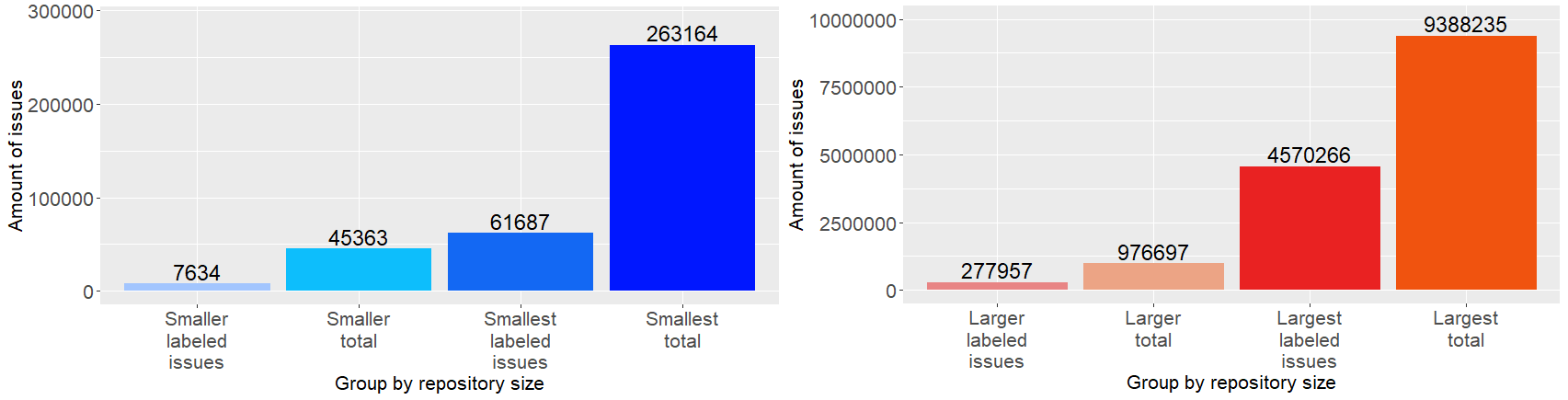}
\caption{Smaller, Smallest, Larger and Lagest labeled issues and total amount of issues.} \label{fig_labeled_small_smallest_larger_lagest}
\end{figure}

After performing the correlation, we analyzed each case in particular.
By splitting the repositories into groups, we could identify outliers in our study and evaluate how repositories with a similar number of issues behave when labeling. As a result, we found that The Largest group had a labeling ratio of 48,68\%, representing 4,570,266 labeled issues. The Larger group labeled 28,45\% of their issues or 277,957 ones. In the Smaller group, 16.82\%, or 7,634 issues, were labeled out of its total. Last, in The Smallest group, 23.44\% or 61,687 issues were labeled. Figure \ref{fig_labeled_small_smallest_larger_lagest} shows such results.

Even with more significant adherence from the repositories to the labeling activity, we could observe that only part of the issues is labeled. As Figures \ref{fig_labeled_small_smallest_larger_lagest} and \ref{fig_labeled_small_smallest_larger_lagest} show, we found out that repositories belonging to the upper median groups (Largest and Larger) are more used to label their issues. We can conclude that the more issues the repository has, the more issues need to be labeled by the contributors. On the other hand, smaller repositories tend to leave the labeling activity aside and focus on resolving issues without this practice.



%% file: Sections/5_Discussions.tex
\section{Discussion}

The analysis of the repositories allowed us to find critical information about the issue labeling activity in the Github repositories. Firstly, we found that the labeling activity is very characteristic in the Github repositories we analyzed, as most of them label at least one issue. We observed that 10,458 (78.75\%) repositories label their issues, while 2,822 (21.25\%) did not label any issues.

Larger repositories prioritize and label many issues. This is likely due to an organizational structure that encourages this type of practice to resolve issues. On the other hand, we observed that small repositories, usually encompassing few issues, do not label their issues.



Another fact observed in repositories that employ labeling activity, is that not all issues are labeled. Considering the set of repositories that label their issue, we found a number of 4,917,544 (46.07\%) labeled issues, while (53,93\%) issue are unlabeled.






Despite having fewer labeled issues, the gain that labeling brings to community engagement and communication is enormous. Our study found that the labeled issues are indeed more striking in the eyes of contributors, with the most significant number of events happening around them and people interacting through the comments of the issues and subscriptions to follow up on the proposed question. If the contributors knew about the gain that an issue brings in numbers, they would label their issues more.

The contributors who see the labeling activity as a good practice started with the first event already labeling issues. We noticed that this practice often becomes the first or second action to be taken at the time of resolution for those who label. Therefore, there is an importance in this activity in the repositories that apply it.

Still, this study's central message is that the labeling activity brings more benefits than not using it. Even if they are large or smaller repositories, the gain in the engagement of personnel and organization when using these structures is significant in compensation for the non-use of the practice.

%% file: Sections/6_Threats_Validity.tex
\section{Threats to Validity}

\subsection{Internal Validity}

Knowing that repositories that belong to different families and organizations bring particularity and different aspects to each other, we seek to find heterogeneity in the repositories using the Github listing, adhering only to the 100 most popular ones. Thus, repositories that did not have issues, or were less frequent, were removed from our study because they did not have enough data analysis input.

\subsection{External Validity} 


We could not generalize the findings of the labeling activity for platforms other than Github and other repositories of different new topics. However, to encompass a more significant number of repositories, we use precisely the list with the repositories of various Github topics with different properties and sizes to seek a more excellent approximation when generalizing the data to others systems of the platform.

In addition, the time of data collection can also be a limiting factor in our study since the repositories used in this study were mined in a specific time frame, and the issues reported later on were not considered in this analysis. For this, we considered the most significant number of valid issues from our database for the study.

\subsection{Construct Validity}

We had an initial concern about the selection metrics of studied repositories. Therefore, we attempted to carefully employ and analyze correlation, labeling time, and the numbers related to the labels themselves. Besides, we were careful to present the data adequately and without interfering with the results.



\subsection{Conclusion Validity}


Considering our study is data-intensive, data analysis through metrics plays a central role in achieving the results. The way we presented it might influence how we analyzed the findings. To avoid misinterpretations or bias, we performed the search and selected different label and usage data representations in the repositories. We employed well-established methods from the literature to measure and present the study data.





%% file: Sections/7_Related_Work.tex
\section{Related Work}

Bissyandé et al. \cite{Bissyande:ISSRE:2013} presented a study that aimed to understand issues and software project activities, including an overview of ten popular tags used in reported issues classified by issue type: bug report and feature request. Also, the study shows a shortlist of all labels and analyzes only the labels bug and enhancement. On the other hand, our study sought to analyze a set of mined repositories, regardless of specific label, the representativeness of the labeling activity in the repositories, including different sizes, the impact labels have on engagement, and issues resolution.

Social and technical aspects were the focus of Treude and Storey \cite{Treude:JournalIEEETransactions:2019}, who studied the creation and behavior of tags on the Jazz platform in an industrial environment, obtaining relevant data on the use of labels in the industrial development domain. Our study analyzes other aspects of a heterogeneous number of open-source repositories, such as the overall impact and engagement in different repositories of different sizes and domains, whose dynamics of use change significantly.

Cabot et al. \cite{Cabot:SANER:2015} explored the use and influence of labels on Github repositories and issues using the GHTorrent \cite{Gousios:MSR:2013} database. The study described the number of labels, the most used ones, and their influence on the project through a correlation between time to solve and issue age. The study also created clusters of types of labels, called label families, and classified them. Our study provides a more in-depth analysis of the representativeness of the labeling activity in different types of repositories. In addition, we considered other metrics to measure the engagement, such as the number of subscriptions to the issue, followers, and comments, and to analyze the level of importance of the labeling at the time of issue resolution getting the labeling moment as a priority method.

Alonso-Abad et al. \cite{Alonso:JournalProgressArtificialIntelligence:2019} studied the prediction of issue labeling using data from seven Github repositories in the GHtorrent database. They presented a list of the leading labels used by these repositories. The study focused on labeling prediction and showed the performance of this practice with text mining techniques and the possibility of indicating closest labels. The focus of the work is on label prediction and performance as analyzing the best techniques for predicting those labels, similar to automatic labeling, unlike our study, which involves an exploratory analysis of the impacts that labels bring to the repository on engagement, repository size, and other aspects.

%% file: Sections/8_Conclusions.tex
\section{Conclusion}

Maintenance and evolution in large-scale systems and communities are among the most significant challenges that open-source systems development could encounter. A platform that provides a good work and communication environment can be decisive for maintaining these communities. Indeed, project settings for more significant community integration have improved significantly. We could state that, in the past, the issues only encompassed their title and description. Modern issue trackers have employed more robust structures, comprising several options for issue data personalization. Among these, labeling has come to assist the developer in solving issues and keeping the gear running more strategically.

With this emerging new use of labels, our study sought to shed light on this activity performed in the latest versions of repositories, based on Github, and show that they are not mere functionalities present on the platform that they offer a rich feature to support developers. Therefore, our study mined 10,673,459 issues from 13,280 repositories of 180 Github featured topics and found that about 78.75\% of the repositories label their issues, with more adherence from those repositories that contain big numbers of issues. The labeling practice is essential and deemed the first step in the issue resolution process in  65.91\% of the first events. Issues with labels draw more attention and impact by collecting more subscribers, assigns, and comments, helping to engage contributors to the resolution.

As future work, we want to study more about the labeling activity and its different types of occurring labels in the repositories and the influence that certain domains have on creating specific labels. We also intend to do a joint analysis of contributors to investigate the creation and use of certain types of labels in the repositories. 


\subsubsection*{Acknowledgments.}
This study was financed in part by the Coordenação de Aperfeiçoamento de Pessoal de Nível Superior - Brasil (CAPES) - Finance Code 001, and FAPESB grant JCB0060/2016.